\documentclass[twocolumn,english,superscriptaddress,showpacs,prl]{revtex4}%
\usepackage[T1]{fontenc}
\usepackage[latin1]{inputenc}
\usepackage{amsmath}
\usepackage{amssymb}
\usepackage{graphicx}
\usepackage{graphicx}
\usepackage{amsfonts}
\usepackage{babel}%
\begin{document}
\author{Alexey A. Kovalev}
\affiliation{Kavli Institute of NanoScience, Delft University of Technology, 2628 CJ Delft,
The Netherlands}
\author{Gerrit E. W. Bauer}
\affiliation{Kavli Institute of NanoScience, Delft University of Technology, 2628 CJ Delft,
The Netherlands}
\author{Arne Brataas}
\affiliation{Department of Physics, Norwegian University of Science and Technology, N-7491
Trondheim, Norway}
\title{Nano-mechanical magnetization reversal}

\begin{abstract}
The dynamics of the ferromagnetic order parameter in thin magnetic films is
strongly affected by the magnetomechanical coupling at certain resonance
frequencies. By solving the equation of motion of the coupled mechanical and
magnetic degrees of freedom we show that the magnetic-field induced
magnetization switching can be strongly accelerated by the lattice and
illustrate the possibility of magnetization reversal by mechanical actuation.

\end{abstract}
\date{\today{}}
\maketitle

The dynamics of the order parameter of small magnetic clusters and films is a
basic problem of condensed matter physics with considerable potential for
technological applications \cite{SpinDyn}. The magnetic field induced reversal
in such systems is perhaps the most active field of research. In clusters as
small as 1000 atoms the magnetization carries out a coherent motion according
to the Stoner-Wohlfarth model \cite{Cluster}. In small magnetic wires, on the
other hand, magnetization reversal is achieved by domain walls traversing the
sample \cite{DW}. With decreasing size of magnetic memory cells the
fundamental limits to the speed and energy dissipation of the magnetization
switching are important issues. Ingenious mechanisms like the so-called
precessional switching \cite{Gerrits:nat02} in which the magnetization vector
traces straight paths on the unit sphere might come close to the optimum in
both respects, but the magnetic fields cannot be strongly localized, its
creation therefore wastes energy. An alternative is the current-induced
spin-transfer torque \cite{Slonc}, that can switch magnetic layers
\cite{CIMR,Valet} as well as move domain walls \cite{Ono:DW}. Completely
different switching strategies, \textit{e.g.} using antiferromagnets
\cite{Kimel:nat04}, attract interest as well.

Advances in fabrication and detection push the fundamental-mode frequencies of
nanodevices up to the GHz range \cite{Cleland,Roukes:nat03}. Nanomechanics and
nanomagnetism come together in magnetic resonance force microscopy that has
already reached single-spin sensitivity \cite{Rugar:nat04}. It has been
suggested that mechanical oscillation can be used to force coherent motions on
uncoupled nuclear spins \cite{Bargatin:prl03}. Here we propose employing the
resonant coupling between magnetic and mechanical degrees of freedom, studied
before only in the limit of small magnetization oscillations
\cite{Kovalev:apl03}, to accelerate magnetization reversal and suggest a
mechanism to switch magnetization by mechanical actuation alone, \textit{i.e.}
without an applied magnetic field. To this end we solve the strongly
non-linear problem in the limit of weak damping and energy transfer
analytically. Comparison with numerical simulations indicates that the
solutions are robust beyond their formal regime of applicability.

We consider a small dielectric cantilever with a single-domain ferromagnetic
layer deposited on its far end (see Fig. 1) in the presence of an external
field $\mathbf{H}_{0}$.\begin{figure}[ptb]
\label{fig1}
\centerline{\includegraphics[  width=5cm]{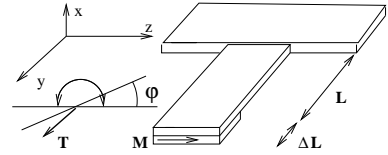}}\caption{A
nano-magneto-mechanical cantilever supporting magneto-vibrational modes. On a
dielectric substrate (such as Si) a single-domain ferromagnetic film is
deposited at the free end.}%
\end{figure}

The dynamics of the magnetization $\mathbf{M}$ is well described by the
Landau-Lifshitz-Gilbert equation \cite{Gilbert:pr55}:%
\begin{equation}
\frac{d\mathbf{M}}{dt}=-\gamma\mathbf{M}\times\mathbf{H}_{\mathrm{eff}}%
+\alpha\mathbf{M}\times\left(  \frac{d\mathbf{M}}{dt}\right)  _{\text{cant}},
\label{LL+current}%
\end{equation}
where $\gamma$ denotes the gyromagnetic ratio. The phenomenological Gilbert
constant is typically $\alpha\leq0.01$, and the derivative $\left(
\frac{d\mathbf{M}}{dt}\right)  _{\text{cant}}=\frac{d\mathbf{M}}{dt}%
+\frac{d\mathbf{\varphi}}{dt}(-M_{\mathrm{z}}\mathbf{x}+M_{\mathrm{x}%
}\mathbf{z})$ is taken in the reference system of the cantilever (see Fig. 1).
For small $\varphi=\varphi(L)$, where $\varphi(y)$ is the torsion angle at
position $y$ of the cantilever, the effective field is $\mathbf{H}%
_{\mathrm{eff}}=(\nu M_{\mathrm{z}}\varphi-\nu M_{\mathrm{x}})\mathbf{x}+\nu
M_{\mathrm{x}}\varphi\mathbf{z}+\mathbf{H_{0}}$, where $\nu$ describes the
demagnetizing dipolar field ($\nu\simeq4\pi$ for our geometry). The coupling
originates here from the demagnetizing field and since the crystal anisotropy
field can be small \textit{(e.g.} in permalloy) it is disregarded
initially.\begin{figure}[ptb]
\label{fig2}
\centerline{\includegraphics[  width=6cm]{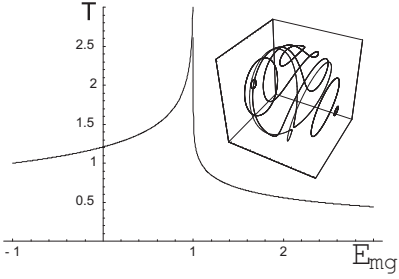}}\caption{The periods
$T_{1}$ ($E_{mg}<1$) and $T_{2}$ ($E_{mg}>1$) of the magnetization dynamics in
Eqs. (\ref{T1/2}) in units of $2\pi/\left(  \gamma\sqrt{(H_{0}+\nu M)H_{0}%
}\right)  $ as a function of energy in units of $H_{0}M$. The inset shows a
plot of typical trajectories at different energies on the unit sphere ($\nu M=10 H_{0}$).}%
\end{figure}

Without coupling and Gilbert damping the dynamics of the magnetic subsystem
can be solved analytically \cite{Serpico:jap03}, leading to the trajectories
and oscillation periods depicted in Fig. 2. The period of the motion can be
expressed by elliptic integrals $K$ as \begin{equation}
\begin{array}{cc}
T_{1}=4\sqrt{2}K\left(p_{-}/p_{+}\right)/\left(\gamma\sqrt{p_{+}}\right)\, ; & E<MH_{0},\\
T_{2}=2\sqrt{2}K\left(1-p_{+}/p_{-}\right)/\left(\gamma\sqrt{-p_{-}}\right)\, ; & E>MH_{0},\\
p_{\pm}=H_{0}^{2}-\nu E\pm H_{0}\sqrt{H_{0}^{2}-2\nu E+\nu^{2}M^{2}} \,, & \end{array}\label{T1/2}\end{equation}
where $E_{mg}=-H_{0}M+ \nu M_{x}^{2}/2$. The strong variation of the periodicity has important consequence for the
coupling to the lattice as explained below.

The equation of motion of the cantilever is \cite{Landau:59}:%
\begin{equation}
C\frac{\partial^{2}\varphi}{\partial y^{2}}=\rho I\frac{\partial^{2}\varphi
}{\partial t^{2}}+2\beta\rho I\frac{\partial\varphi}{\partial t},
\label{wave equation}%
\end{equation}
where $C$ is an elastic constant defined by the shape and material of the
cantilever ($C=\frac{1}{3}\mu da^{3}$ for a plate with thickness $a$ much
smaller than width $d$ and $\mu$ is the Lamé constant), $I=\int(z^{2}%
+x^{2})dzdx\simeq ad^{3}/12$ is the moment of inertia of the cross-section
about its center of mass, $\rho$ the mass density, and $\beta$ a
phenomenological damping constant related to the quality factor $Q$ at the
resonance frequency $\omega_{e}$ as $Q=\omega_{e}/(2\beta)$ (at 1 GHz
$Q\sim500$ \cite{Roukes:nat03}). Note that $\omega_{e}$ can be also a higher
harmonic resonance frequency in what follows. The clamping boundary condition
is $\varphi|_{y=0}=0$. The conservation law for the mechanical angular
momentum $\mathbf{V}^{\mathrm{el}}(y)$ for a thin slice at point $y\in\left\{
0,L\right\}  $ (without magnetic overlayer) $d\mathbf{V}^{\mathrm{el}}\left(
y\right)  /dt=\mathbf{T}\left(  y\right)  ,\,\,$where the torque
$\mathbf{T}(y)$ flowing into the slice, is modified by the coupling to the
magnet in a region $y\in\left\{  L,L+\Delta L\right\}  $ ($\Delta L$ is the
length of the cantilever covered by the magnetic layer) as:%
\begin{equation}
\frac{d}{dt}\left(  \mathbf{V}^{\mathrm{el}}\left(  L\right)  +(-\frac
{1}{\gamma})\mathbf{M}\left(  L\right)  V\right)  =\mathbf{T}\left(  L\right)
+\mathbf{T}_{\text{field}} \, , \label{motion}%
\end{equation}
where $\mathbf{T}_{\text{field}}=\mathbf{M}\times\mathbf{H}_{0}$, $V$ volume
of the magnet and $\mathbf{T}|_{y}=-C\tau\left(  y\right)  $ is the torque
flowing through the cantilever at point $y$ ($\tau=\partial\varphi(y)/dy$).
When $\Delta L\ll L,$ internal strains in the magnetic section may be
disregarded. The magnetovibrational coupling can then be treated as a boundary
condition to the mechanical problem \cite{Kovalev:apl03}, which is expressed
as the torque $C\tau|_{y=L}$ exerted by the magnetization on the edge of the
cantilever:
\begin{equation}
C\tau|_{y=L}=\frac{1}{\gamma}\left(  \frac{d\mathbf{M}}{dt}+\gamma
\mathbf{M}\times\mathbf{H}_{0}\right)  |_{y} \, . \label{bound}%
\end{equation}
The effect of the coupling between Eqs. (\ref{LL+current}) and
(\ref{wave equation}) may thus be summarized as%
\begin{equation}%
\begin{array}
[c]{c}%
\frac{\partial\varphi}{\partial y}|_{y=L}=\sqrt{\nu M/H_{0}}\frac{\pi\varphi_{0}^{2}}{\left(
cM\right)}  \left(  \frac{dM_{\text{y}}}{dt}+\gamma\mathbf{M}\times
\mathbf{H}_{0}\right)\, , \\
\mathbf{H}_{\mathrm{eff}}=(\nu M_{\mathrm{z}}\varphi-\nu M_{\mathrm{x}%
})\mathbf{x}+\nu M_{\mathrm{x}}\varphi\mathbf{z}+\mathbf{H}_{0} \, ,
\end{array}
\label{MME}%
\end{equation}
where we introduce the parameter $\varphi_{0}^{2}=MV\sqrt{H_{0}/\left(  \nu
M\right)  }/\left(  \gamma2\rho IL\omega_{e}\right)  $, whose physical meaning
is explained below. In the absence of damping, the above system of equations
can be obtained as well from the free energy:%
\begin{equation}
F=V(-\mathbf{M}\mathbf{H}_{0}+\frac{\nu}{2}M_{\mathrm{x}}^{2}+\nu
M_{\mathrm{x}}M_{\mathrm{z}}\varphi(L))+\frac{C}{2}\int_{0}^{L}\left(
\frac{\partial\varphi}{\partial x}\right)  ^{2}dx. \label{Free Energy}%
\end{equation}

We first address the coupling strength \cite{max} of a system described by Eq.
(\ref{Free Energy}). Consider the two subsystems oscillating at common
frequency $\omega$. The total mechanical energy is then $E_{me}=\rho
IL\omega^{2}\varphi_{0}^{2},$ where $\varphi_{0}$ is the maximal angle of the
torsional motion (that will turn out to be identical to the parameter
introduced below Eq. (\ref{MME})). By equipartition this energy should be of
the order of the magnetic energy $E_{mg}=MVH_{0}$. The maximal angle would
correspond to the mechanical motion induced by full transfer of the magnetic
energy to the lattice. By equalizing those energies we find an estimate for
the maximal angle of torsion $\varphi_{0}=\sqrt{MVH_{0}/\left(  \rho
IL\omega^{2}\right)  },$ which at resonance is identical to the parameter
introduced above. The coupling between the subsystems can be measured by the
distribution of an applied external torque (\textit{e.g}. applied by a
magnetic field) over the two subsystems. The total angular momentum flow into
the magnetic subsystem by the effective magnetic field is $(MV/\gamma)\omega,$
whereas that corresponding to the mechanical subsystem at the same frequency
is $\left(  \rho IL\omega\varphi_{0}\right)  \omega.$ Their ratio is
$\varphi_{0}\sqrt{\nu M/H_{0}}$. The maximum angle $\varphi_{0}$ derived above
is therefore also a measure of the coupling between the magnetic and
mechanical subsystems. This estimate is consistent with the splitting of
polariton modes at resonance of $G=\varphi_{0}^{2}\omega^{2}\nu M/H_{0}$
\cite{Kovalev:apl03}. An estimate for a cantilever with $\rho
=2330\,\,\mathrm{kg/m}^{3}$ (Si) and $d=1$ $\mathrm{\mu m}$ leads to
$\varphi_{0}^{2}\nu M/H_{0}\sim M/\left(  \gamma^{2}\rho d^{2}H_{0}\right)
\sim10^{-3}$. Decreasing $d$, $\rho$, $H_{0}$ or $1/M$ is beneficial for the coupling.

Magnetization reversal by a magnetic field in the coupling regime can be
realized even without any damping by transferring magnetic energy into the
mechanical system. Since we find that $\varphi_{0}\ll1$ for realistic
parameters, the subsystems undergo many precessions/oscillations before the
switching is completed. The switching is then associated with a slow time
scale corresponding to the global motion governed by the coupling or a
(reintroduced) weak damping relative to a fast time scale characterized by the
Larmor frequency. The equation of motion for the slow dynamics (the envelope
functions) can be derived by averaging over the rapid oscillations. To this
end we substitute Eq. (\ref{MME}), linearized in the small parameters $\alpha
$, $\beta$ and $\varphi_{0}$, into the equations for the mechanical and the
magnetic energies:%
\begin{equation}%
\begin{array}
[c]{c}%
\frac{d}{dt}E_{me}=-2\beta E_{me}+C\tau|_{x=L}\frac{d\varphi}{dt}|_{x=L},\\
\frac{d}{dt}E_{mg}=-H_{0}\dot{M}_{z}+\nu M_{\mathrm{x}}\dot{M}_{x} \, .
\end{array}
\label{Eng}%
\end{equation}
We focus in the following on the regime $H_{0}\ll\nu M$, which usually holds
for thin films and not too strong fields, in which the magnetization motion is
elliptical with long axis in the plane and small $M_{x}$ even for larger
precession cones. Disregarding terms containing higher powers of $M_{x}$ and
averaging over one period as indicated by $\left\langle ...\right\rangle $%
\begin{equation}%
\begin{array}
[c]{c}%
\left\langle \frac{dE_{me}}{dt}\right\rangle +\left\langle 2\beta
E_{me}\right\rangle =-V\nu\left\langle M_{z}M_{x}\dot{\varphi}\right\rangle
,\\
\left\langle \frac{dE_{mg}}{dt}\right\rangle +\left\langle \alpha%
\nu^{2} M M_{x}^2\right\rangle =\nu\left\langle M_{z}\dot{M}%
_{x}\varphi\right\rangle .
\end{array}
\label{Eng1}%
\end{equation}
By adiabatic shaping of time-dependent magnetic fields we can keep the two
subsystems at resonance at all times. The slow dynamics $\varphi(L,t)\sim
A(t)e^{i\left(  \omega+\pi/2\right)  t}$ and $M_{x}\sim W(t)e^{i\omega t}$ in
time domain is then governed by the equation:%
\begin{equation}%
\begin{array}
[c]{c}%
\dot{A}+\beta A=-\frac{\varphi_{0}^{2}\gamma\nu}{H_{0}M}\sqrt{\nu M/H_{0}%
}(-MH_{0}+\frac{\nu W^{2}}{4})W \, ,\\
\dot{W}+\alpha\sqrt{\nu M/H_{0}}\omega W/2=\frac{\omega}{H_{0}}(-MH_{0}%
+\frac{\nu W^{2}}{4})A \, ,
\end{array}
\label{oscillations1}%
\end{equation}
$\varphi(L,t)\sim\widetilde{A}(t)e^{i\omega t}$ and $M_{x}\sim\widetilde
{W}(t)e^{i(\omega-\pi/2)t}$ corresponding to $\pi/2$-shifted harmonics are
also solutions, and the initial conditions determine the linear combination of
two envelope functions, \textit{i.e}. the beating pattern of two hybridized
polariton modes \cite{Kovalev:apl03}. When initially all energy is stored in
one degree of freedom, $M_{x}$ is $\pi/2$ shifted from $\varphi(L,t)$ and
$A_{2}(t)=0$. Eqs. (\ref{oscillations1}) describe a (damped) harmonic
oscillator with frequency $\omega\varphi_{0}\sqrt{\nu M/H_{0}}=G\ll\omega$
when $\nu W^{2}/4\ll MH_{0}$. Such oscillatory behavior persists for general
angles (except for motion with very large angle cones close to the
antiparallel configuration at which the frequency is reduced by up to a factor
1/2). This is illustrated by Fig. 3 which shows a numerical simulation of Eq.
(\ref{MME}) for an undamped system excited at $t=0$ by a magnetic field
$\vec{H}_{0}$ at an angle $2\pi/3$ with the initial magnetization. The number
of periods necessary to transfer all energy from one subsystem to the other is
therefore given by $\sim1/\left(  4\varphi_{0}\sqrt{\nu M/H_{0}}\right)  $.
Eq. (\ref{oscillations1}) also shows that for damping constants $\alpha
>\varphi_{0}/\pi$ or $\beta/\omega>\varphi_{0}\sqrt{\nu M/H_{0}}/\pi$ the
beating is suppressed. \begin{figure}[ptb]
\label{fig3}
\centerline{\includegraphics[  width=6cm]{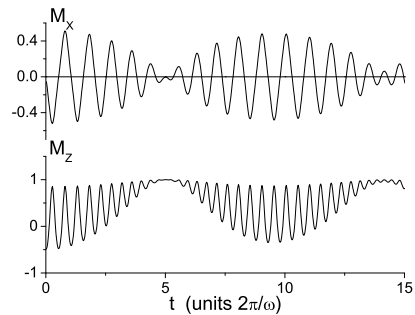}}\caption{Time-dependent
response of the magnetomechanical system to an external magnetic field
switched on at $t=0$, in the absence of dissipation. Plotted are the $x-$ and
$z-$components of the magnetization ($\nu M=10 H_{0}$). }%
\end{figure}\begin{figure}[ptbptb]
\label{fig4}
\centerline{\includegraphics[  width=7cm]{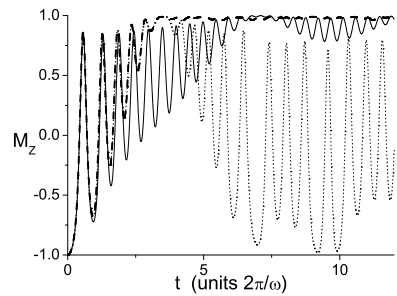}}\caption{Magnetization
reversal assisted by the mechanical coupling. In the dashed line the
external magnetic field is reduced at moment $t=3.5$ to half of its initial
value (at $\alpha=\beta=0)$. In the dotted and solid lines the field is
kept constant. The solid line illustrates the effect of a mechanical
damping of $\,\,\beta=0.04$  ($\nu M=10 H_{0}$).}%
\end{figure}

Fig. 3 illustrates that the mechanical system absorbs energy from the magnetic
subsystem and gives it back repeatedly in terms of violent oscillations that
are modulated by an envelope function on the time scale derived above. When
the envelope function vanishes the magnetization is reversed and the systems
seems to be at rest. However, since the energy is not dissipated, the
momentarily silence is deceptive, and the beating pattern repeats. An
efficient coupling requires that the frequencies of the subsystems are close
to each other at each configuration, which was achieved in the simulation by
the adiabatic modulation of the magnetic field $H_{0}$ according to Fig. 2.
However, the reversal process is robust; an estimate from Eqs. (\ref{Eng1})
for the necessary proximity of the resonant frequencies of the mechanical and
the magnetic subsystems is $\Delta\omega\sim\left(  \nu M/H_{0}\right)
^{1/2}\varphi_{0}\omega$. In that case the above estimates still hold.
\begin{figure}[ptb]
\label{fig5}
\centerline{\includegraphics[  width=7cm]{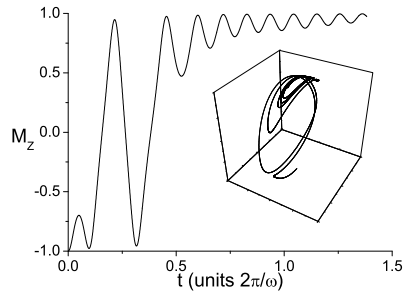}}\caption{Magnetization
switching by a pulsed actuation of a damped cantilever without external field.
Inset shows the corresponding magnetization trajectory on the unit sphere.}%
\end{figure}

Let us now consider the reversal from the state $M_{z}=-1$ by pulsed actuation
of the mechanical system with an energy less than $H_{0}M$ by twisting the
cantilever abruptly at $t=0$. The mechanical actuation is not essential here,
but it helps the magnetization to quickly escape the instable equilibrium
point. We can suppress the backflow of energy by sufficiently damping the
mechanical subsystem as illustrated in Fig. 4 for $\beta\sim0.04$ with a
vanishing Gilbert damping, $\alpha=0$. Alternatively we may detune the
external magnetic field out of the resonance after the first reversal. We
observe that even without intrinsic damping the unwanted \textquotedblleft
ringing\textquotedblright\ after the switching is strongly suppressed. What
happens here is that starting with all energy stored in the magnetic degree of
freedom, the system is brought into resonance only for the time of one beating. All
energy is then irreversibly transferred to the mechanical degree of freedom.

Finally we propose a non-resonant mechanical reversal scheme analogous to
\textquotedblleft precessional\textquotedblright\ switching
\cite{Gerrits:nat02}. The effective field $\mathbf{H}_{\mathrm{eff}}$ (see Eq.
(\ref{MME})) has a component perpendicular to the plane of the film $H_{x}\sim
M\varphi$. Under a sudden mechanical twist this component acts like a
transverse magnetic pulse about which the precessing develops. The mechanical
actuation should be fast, on a scale $(\gamma\varphi\nu M)^{-1},$ but there
are no restrictions on the mechanical frequency now. We integrate this system
numerically for a strongly damped cantilever $\beta/\omega\sim1$ and a twist
of $\varphi=0.2$. In Fig. 5 we plot the reaction of $M_{z}$, reintroducing an
easy axis anisotropy described by $DM_{z}\mathbf{z}$ ($D=0.05$, $\alpha=0.01$)
and in the absence of an external field. As in the case of the precessional
switching, the switching time can be minimized by a more careful adjustment of
the mechanical pulse in order to realize the optimum \textquotedblleft
ballistic\textquotedblleft\ path between $M_{z}=\pm1$.

Summarizing, we investigated the non-linear dynamics of coupled magnetic and
mechanical fields for a cantilever with a ferromagnetic tip. Conventional
magnetic-field induced reversal can be accelerated by finding new materials
with higher dissipation, \textit{i.e}. Gilbert damping constant $\alpha$. We
propose here three strategies for fast magnetization reversal and suppressed
\textquotedblleft ringing\textquotedblright\ using conventional magnets but
opening new dissipation channels by i) making use of the additional mechanical
damping, ii) shaping the external magnetic field pulses, thus quickly
channeling-off magnetic energy when damping is weak. Furthermore, we propose
iii) a precessional reversal due to a mechanically generated out-of-plane
demagnetizing field without applied magnetic fields.

The experimental realization will be a challenge since the cantilever has to
work at very high frequencies (note that coupling to higher harmonics may be a
partial solution) or has to be actuated fast. The magnetic film should be
small enough to form a single domain. With the present pace of progress in
nanomechanics and nanomagnetism, we are optimistic that these conditions can
be met in the not too far future.

This work has been supported by the Dutch FOM Foundation and the Research
Council of Norway, NANOMAT Grants. No. 158518/143 and 158547/431. We thank
Yaroslav Tserkovnyak and Oleg Jouravlev for helpful discussions.


\begin{thebibliography}{99}
\bibitem {max}In principle a ferromagnetic cantilever would achieve maximal
coupling when the volume of the magnet equals the volume of the cantilever.
Unfortunately such a system is more difficult to analyze and is therefore not
considered here.
\end{thebibliography}
\end{document}